\documentclass[aps,prl,twocolumn,epsfig,floats]{revtex4}
\usepackage{graphicx,epsf}

\begin{document}
%\documentstyle[aps,epsf,preprint]{revtex}

%\everymath{\rm}
%\everydisplay{\rm}
%\renewcommand{\baselinestretch}{1.5}
%\setlength{\textwidth}{6.25in}

\title{Spin Order accompanying Loop-Current Order in Cuprates}
\author{Vivek Aji and C. M. Varma}
\address{Physics Department, University of California,
Riverside, CA 92507}
\begin{abstract}

The theory of the long range order of orbital current loops in the pseudogap phase is generalized to include the effects of
spin-orbit scattering. It is shown by symmetry arguments as well as by microscopic calculation that a specific in-plane
spin-order must necessarily accompany the loop-current order. The microscopic theory also allows an estimate of the
magnitude of the ordered spin-moment. Exchange coupling between the generated spins
further modifies the in-plane direction of the
spin moments. The structure and form factor for the spin and orbital
moments combined with the induced spin order is consistent with the direction of moments deduced from
polarization analysis in the  neutron scattering experiment.


\end{abstract}

\maketitle

\section{Introduction}

\begin{figure}
  \includegraphics[width=0.7\columnwidth]{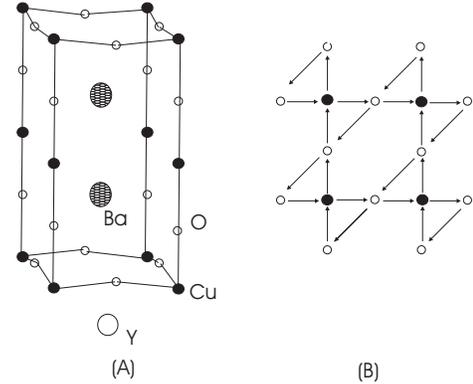}
  \caption{(A)Crystal structure of YBCO and (B) the Current pattern in the observed time reversal violating states}
  \label{cp}
\end{figure}

In all underdoped cuprates thermodynamic, transport, and
spectroscopic measurements reveal the formation of a new state of
matter which is commonly referred to as the "pseudogap" state. Any
credible theory of cuprates  must specify the nature of this state.
One of the theories proposed predicts that this phase breaks time
reversal through ordered current loops in o-cu-o plaquettes without
breaking translational symmetry. \cite{CMV1,CMV2,SV}. The
fluctuations about this state are such that there is no
specific-heat singularity at the transition \cite{Aji1}. This
removes the major difficulty in regarding that the pseudogap state
represents a broken symmetry. The unit cell for YBCO is shown in
fig.(\ref{cp}A). The predicted loop-current order in the copper
oxide planes has the pattern shown in fig.(\ref{cp}B)for one of the
four possible domains.  Evidence for such a state was obtained from
ARPES using circularly polarized light \cite{AK} in BISCCO, and more
directly by recent polarized neutron scattering diffraction in YBCO
\cite{FAQ}.

While the spatial symmetry of moments of fig.(\ref{cp}B) is borne
out by the neutron experiments, the direction of the magnetic
moments is not consistent with the predictions. The orbital moments
should be normal to the o-cu-o plaquettes. The plaquettes are not
co-planar with the two-dimensional cu-planes due to the buckling of
the planes (see fig.\ref{cp}(A)); for YBCO in which the neutron
scattering experiments were done the nearest neighbor o-cu-o
plaquettes make an angle of about $7^{\circ}$ with respect to the
Cu-planes. Therefore a tilting of only about $7^{\circ}$ of the
moments with respect to the normal to the Cu-planes is expected.
However this angle has been deduced to be $45\pm 20^{\circ}$
\cite{FAQ}.

The purpose of this paper is to resolve this matter. The basic
physical point we draw on \cite{CJZ}  is that spin orbit interaction
can lead to spin ferromagnetism in states with orbital currents
\cite{ftwu}. We present general symmetry arguments supporting this
and calculate microscopically the nature of spin order in YBCO for
states of the symmetry consistent with the observations \cite{FAQ}.
The magnitude of the spin-moment is estimated to be only $10-20$\% of
the orbital moment. But the neutron scattering intensity depends on
the spatial distribution or the form-factor of the moments besides their
magnitudes. For the $[011]$ Bragg peak studied in experiments, we find that
the spin form factor is significantly larger than the orbital
current form factor because the latter are spread out more inside a unit-cell than the former. The existing experimental results may thus
reconciled with the theory but a definitive confirmation awaits the measurement of the form-factors in experiments.

The direction of the ordered in plane spin-moments is affected also
by the exchange interaction between the moments. We present a rough
order of magnitude for this effect  which suggests that the spin
order may be quite complicated. Reliable theoretical estimates on
the actual spin-order are very difficult to make at this point.
However, there are some general features of the results which are
expected to be robust. The details of the magnetic order suggested
as possible here can only be resolved in experiments which have
greater accuracy than the one performed to date. A companion to this
paper contains the details of the experimental results which were
published as a short report earlier as well as an analysis of the
data applying the ideas in this paper \cite{YS, FQ}.

\section{Symmetry Considerations}

We can deduce through arguments using the symmetry of the crystal
structure, and of the orbital order parameter, Fig.~\ref{cp}, that
spin-orbit coupling must induce an in-plane order of the spins.
First we present an argument by constructing a free-energy invariant
which is a product of the spin-orbit tensor, the orbital order and
the possible spin-order. We will  show that the same conclusions can
also be obtained from general arguments patterned after those due to
Dzyaloshinskii \cite{ID}. These arguments allow one to deduce the
symmetry of the spin order parameter but the physical basis of these
general arguments requires a microscopic theory which also allows us
to obtain the magnitude of the effective moments.

The crystal structure of YBCO is shown in fig.\ref{cp}(A). Notice
that the Oxygen atoms are displaced from the plane formed by the
Copper atoms; the copper oxide plane is not flat but dimpled. Cu's do not sit at a point of {\it Inversion}. This
leads to a linear coupling between the spin and orbital degrees of
freedom.

Consider first the symmetries of the crystal structure. Due
to the buckling of the Oxygens, the crystal structure in
fig.\ref{cp}(A)

\noindent (i) breaks inversion symmetry through Copper and Oxygen

\noindent (ii) but preserves reflection symmetry about ${\bf \hat{x}},
{\bf \hat{y}}, {\bf \hat{x}}+{\bf \hat{y}}$ and ${\bf \hat{x}}-{\bf
\hat{y}}$.

\noindent The spin-orbit coupling tensor ${\bf \Lambda}$, which
couples linearly the spin and the momentum of the electrons,
respects these symmetries. Let us now look at the symmetries of the
current loop order. The order parameter ${\bf M_O}$, corresponding
to the domain in fig.\ref{cp}(B),

\noindent (i) breaks inversion symmetry through Copper and Oxygen

\noindent (ii) preserves the reflection symmetry about
($\hat{\textbf{x}}-\hat{\textbf{y}}$)

\noindent (iii) breaks reflection symmetries about (${\bf \hat{x}}$,
${\bf \hat{y}}$, ${\bf \hat{x}}+{\bf \hat{y}}$)

  Let $M_S$ specify the distribution and
direction of possible spin-order which obeys the translational
symmetry of the crystal and other symmetries so that it {\it must} accompany the orbital order. An invariant  term in the
free-energy of the form

\begin{equation}
f_{so} = \Lambda^{\alpha\beta}M_{O}^{\alpha}M_{S}^{\beta} +
{M_{S}^{2}\over \chi}
\label{fso}
\end{equation}

\noindent must then  exist. Here $\chi$ is the spin susceptibility
for the order specified by $ M_{S} $. Given  eqn.\ref{fso},
\begin{equation}
 M_{S}^{\alpha} = \chi
\Lambda^{\alpha\beta}M_{O}^{\beta}
\end{equation}

\noindent will be realized.

Consider the symmetries that need to be satisfied
by $\textbf{M}_{S}$. The product of
$\Lambda$ and $M_{O}$ preserves inversion through Copper and Oxygen,
but breaks time-reversal and the one mentioned reflection. Hence
$\textbf{M}_{S}$ must be

\noindent (i) odd under reflection about (${\bf \hat{x}}$, ${\bf
\hat{y}}$, ${\bf \hat{x}}+{\bf \hat{y}}$)

\noindent (ii) even under reflection about
($\hat{\textbf{x}}-\hat{\textbf{y}}$)

\noindent (iii) even under inversion about Copper and Oxygen

\begin{figure}
  \includegraphics[width =0.7\columnwidth]{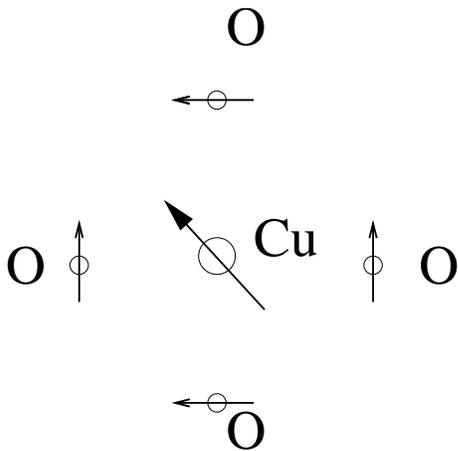}
  \caption{Calculated spin order in YBCO for the domain shown in fig.\ref{cp}(B) in the absence
  of exchange interaction}
  \label{so}
\end{figure}

 A spin-order ${\bf M_S}$ consistent with
these requirements  is  shown for one of the two
cu-o bilayers in fig.\ref{so}. Under inversion through Copper, the
spins remain the same thus satisfying inversion (since spin is an
axial vector). All reflections other than $\textbf{x}=\textbf{y}$
are broken. Since the oxygens layers are below the cu-layers in one
of the bi-layers and above it in the other, the spin-orbit coupling
has opposite sign in the two bilayers. It follows that the direction
of the moments specified by ${\bf M_S}$ is opposite in the two
bi-layers so that the net moment per unit-cell is zero.

In the argument above $\left<\Lambda M_{O}\right>$ acts as a net
magnetic field on the spins and the order is stabilized by the
quadratic term in the free energy. One can also give an argument, which we find a bit more abstract, following
Dzialoshinskii-Moriya (DM)  \cite{ID,TM} for  ${\bf M_S}$.
 In DM, one asks whether an
anti-symmetric interaction between magnetic moments ${\bf M}_A$ and
${\bf M}_B$ of the form

\begin{equation} {\bf D}_{AB}\cdot ({\bf
M}_A\times{\bf M}_B)
\label{DM}
\end{equation}

\noindent has the symmetries of the lattice. The direction of  ${\bf
D}_{AB}$ is specified by the crystal symmetry in relation to the position
A and B of ${\bf M}_A$ and ${\bf M}_B$. The general conditions on
${\bf D}_{AB}$ have been given by Moriya \cite{TM}. We take ${\bf M}_A$
and ${\bf M}_B$ to be the moments at the position of the centroid of
the two triangles with the currents shown in Fig.~\ref{cp}B. Due to
the buckling of the planes, there is no center of inversion in the
vector connecting these two moments. Then ${\bf D}_{AB} \ne 0.$ A mirror
plane perpendicular to AB bisects AB. Then ${\bf D}_{AB}$ must be
parallel to the mirror plane. There exists also twofold rotation
axis perpendicular to AB which passes through the mid-point of AB.
Then ${\bf D}_{AB}$ must be perpendicular to the two-fold axis. Thus
${\bf D}_{AB}$ is along the c-axis and in the mirror plane specified by
its normal ${\bf \hat{x}}-{\bf \hat{y}}$ passing through cu. Given such a ${\bf D}_{AB}$, a
tilt of  ${\bf M}_A$ and ${\bf M}_B$ so that they have a finite in-plane is mandated by the term (\ref{DM}). This is consistent with
the direction of the spin-order deduced from the previous argument
and shown in Fig.~\ref{so}.

An important point to emphasize here is that the symmetry
considerations  do not specify the relative orientation of the spin
on the Copper and Oxygen atoms. The moment in the unit-cell has a component in the  $-\textbf{x}+\textbf{y}$ direction but the relative direction of the
spin- moment on Oxygen and on Copper spin are not specified. The
microscopic calculation in the next section based purely on
spin-orbit scattering provides the result that these moments are parallel. However, once a
spin-moment is generated by the "effective field" provided by
spin-orbit coupling and orbital order, one must consider also the
exchange interaction between them. The actual spin arrangement
depends on the relative magnitude of the exchange interactions to
the "effective fields" for spin order due to orbital order and
spin-orbit coupling. The exchange interactions will be
considered later.

\begin{figure}
  \includegraphics[width =0.7\columnwidth]{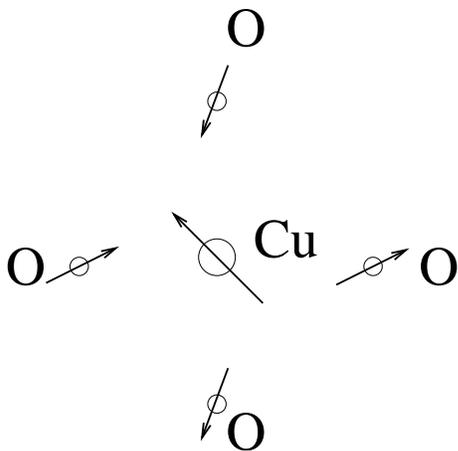}
  \caption{Schematic spin order in YBCO for the domain shown in fig.\ref{cp}(B) in the
  presence of exchange interaction. The spins need not be collinear and the angle
  between the net oxygen spins (summed over all oxygens in an unit cell) and the Copper
  atom is dictated by the relative strength of the exchange and ordering fields.}
  \label{soj}
\end{figure}

\section{Microscopic Model}

The theory of loop-current order is derived from a microscopic model
on the basis of  the Copper $d$ and the two Oxygen $p_{x,y}$
orbitals in each unit cell \cite{CMV2}. The unit cell is shown in
fig.\ref{so}. The Hamiltonian for the system is

\begin{eqnarray}\label{ham1}
H &=& \sum_{i
\alpha}\left[\bar{t}_{pd}d_{i\alpha}^{\dagger}\left(p_{i+x
\alpha}-p_{i-x\alpha}+p_{i+y\alpha}-p_{i-y\alpha}\right)\right. \\
\nonumber &+& \left.
\bar{t}_{pp}\left(p_{i+x\alpha}^{\dagger}-p_{i-x\alpha}^{\dagger}
\right)\left(p_{i+y\alpha}-p_{i-y\alpha}\right) \right] \\
\nonumber &+& V\sum_{i\alpha\beta}
d_{i\alpha}^{\dagger}d_{i\alpha}\left(p_{i+x\beta}^{\dagger}p_{i+x\beta}+p_{i+y\beta}^{\dagger}p_{i+y\beta}\right)
+ c.c.
\end{eqnarray}

\noindent where $\alpha$ and $\beta$ are spin indices, the sum is
over the position of the Copper atoms, $\bar{t}_{pd}$ and
$\bar{t}_{pp}$ are renormalized hopping which include the effects of
the large on site repulsions, and $V$ is the Coloumb repulsion
between the charges on the Copper and Oxygen. The symmetry breaking
is captured by considering the mean field decoupling of the quartic
term. An operator identity for the interaction terms is:

\begin{eqnarray}\label{ans}
d_{i\alpha}^{\dagger}d_{i\alpha}p_{i+x\beta}^{\dagger}p_{i+x\beta}
&=&  {1\over 2}\left(-\left|j_{i,i+x\alpha\beta}\right|^{2} \right.
\nonumber \\ &+&\left.  d_{i\alpha}^{\dagger}d_{i\alpha} +
p_{i+x\beta}^{\dagger}p_{i+x\beta}\right).
\end{eqnarray}
\noindent Here $j_{i,i+x\alpha\beta}$ is the current-tensor between
the sites $i$ and $i+x$.
\begin{equation}
\\j_{i,i+x\alpha\beta} = \imath \left(d_{i\alpha}^{\dagger}p_{i+x\beta}  - p_{i+x\beta}^{\dagger}d_{i\alpha}\right)
\end{equation}

\noindent The mean field ansatz is
\begin{equation}
\sqrt{V} j_{i,i\pm x\alpha\beta} = \pm\sqrt{V} j_{i,i\pm
y\alpha\beta} = \psi \delta_{\alpha\beta} =
R\exp\left(\imath\varphi\right)
\end{equation}

\noindent So an order parameter in terms of ordinary current (and
not spin-current) is sought. Symmetry requires $\varphi = \pi/2 $.

Due to the buckling of the planes there is finite overlap between
the Oxygen $(p_x,p_y)$ and the  Copper $(d_{xz},d_{yz})$ orbitals
respectively. The spin orbit interaction on Copper couples the
$d_{x^{2}-y^{2}}$ orbital with the $d_{xz}$ and $d_{yz}$ orbitals
but such matrix elements are off diagonal in spin. Thus processes
are allowed wherein the electron can hop from the ground state
orbital of the Copper to that of Oxygen and also flip its spin in
the process. The effective hamiltonian generated by such processes
is of the form  \cite{CBT,CRZ}:
\begin{equation}\label{hso}
H_{so} = \sum_{i,\delta} \imath d_{i\alpha}^{\dagger}
\overrightarrow{\lambda}_{i,i+\delta}\cdot
\overrightarrow{\sigma}_{\alpha\beta}p^{x}_{i+\delta \beta} + c.c.,
\end{equation}
\noindent where $\delta = \left(\pm \textbf{x}, \pm
\textbf{y}\right)$, $\lambda_{i,i\pm \textbf{x}} =
\lambda_{0}\widehat{y}$ and $\lambda_{i,i\pm \textbf{y}} =
-\lambda_{0}\widehat{x}$. The coupling constant is given by $
\lambda_{0} =\left|{
\left<d_{x^{2}-y^{2}}\right|\textbf{L}\left|d_{xz}\right>t_{xz}/
\epsilon_{0}}\right|$,  where $\textbf{L}$ is the angular momentum
operator, $t_{xz}$ is the hopping matrix element between the
$d_{xz}$ and $p_{x}$ orbitals and $\epsilon_{0}$ is the energy
difference between $d_{x^{2}-y^{2}}$ and $d_{xz}$ orbitals. The
total Hamiltonian for the system is

\begin{equation}
H  = H_{mf} + H_{so} + H_{ex}
\end{equation}

\noindent We have included the spin exchange term in the
Hamiltonian, $H_{ex}$ which we will discussed later. The Mean Field
Hamiltonian \cite{CMV2} is obtained from from eqn.\ref{ham1} with the mean field
ansatz (eqn.\ref{ans}) made to decouple the quartic interaction.

We first determine the spin state (in the absence of the exchange
coupling discussed below)  and see that it indeed reproduces the results from
general symmetry grounds obtained above. Fourier transforming the Hamiltonian, in
the basis $
\{d_{\textbf{k}\uparrow},d_{\textbf{k}\downarrow},p_{\textbf{x}\textbf{k}\uparrow},p_{\textbf{x}\textbf{k}\downarrow},
p_{\textbf{y}\textbf{k}\downarrow},p_{\textbf{y}\textbf{k}\uparrow}\}$
 $H =$ $ H_{mf}$ $ +H_{SO}$ is given by :
\begin{widetext}
\begin{equation}\label{ham}
H = \left(
  \begin{array}{ccc}
    0 & 2 \imath\left(\bar{t}_{pd}s_{x}\left(\textbf{k}\right)+ R c_{x}\left(\textbf{k}\right)\right)\textbf{I} + \imath\lambda_{0} c_{x} \sigma_{y}  & 2\imath\left(\bar{ t}_{pd}s_{y}\left(\textbf{k}\right)+ R c_{y}\left(\textbf{k}\right)\right)\textbf{I}-\imath\lambda_{0} c_{x}\sigma_{x} \\
    2 \imath\left(\bar{t}_{pd}s_{x}\left(\textbf{k}\right)+ R c_{x}\left(\textbf{k}\right)\right)\textbf{I} + \imath\lambda_{0} c_{x} \sigma_{y} & 0 & 4 \bar{t}_{pp}s_{x}\left(\textbf{k}\right)s_{y}\left(\textbf{k}\right)\textbf{I} \\
    2 \imath\left(\bar{t}_{pd}s_{y}\left(\textbf{k}\right)+ R c_{y}\left(\textbf{k}\right)\right)\textbf{I} - \imath\lambda_{0} c_{x}\sigma_{x} & 4 \bar{t}_{pp}s_{x}\left(\textbf{k}\right)s_{y}\left(\textbf{k}\right)\textbf{I} & 0 \\
  \end{array}
\right)
\end{equation}
\end{widetext}
\noindent where $s_{x,y}\left( \textbf{k}\right) =
\sin\left(k_{x}a/2,k_{y}a/2\right)$, $c_{x,y}\left(
\textbf{k}\right) = \cos\left(k_{x}a/2,k_{y}a/2\right)$,
$\textbf{I}$ is the identity matrix, and $\sigma$'s are the pauli
matrices. Consider first no spin-orbit coupling, i.e.
$\lambda_{0}=0$. This mean field Hamiltonian lead to the
Time-reversal breaking of the loop-current phase with order
parameter $R$,  but for $\lambda=0$ it preserves spin rotational
invariance. The minima of the band is shifted from the $\Gamma$
point corresponding to the fact that the ground state breaks
inversion and the reflection symmetry $-\textbf{x}+\textbf{y}$. The
particular direction of wave-vector picked out by the ground state
depends on the choice of domain.

\section{Spin Order}

\begin{figure}
  \includegraphics[height =2in]{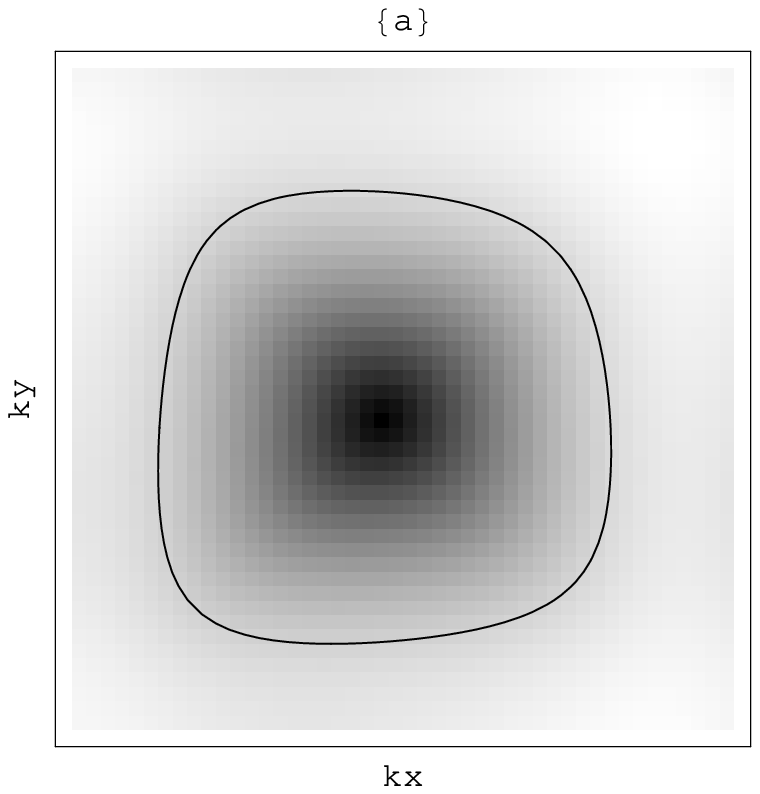}
  \includegraphics[height =2in]{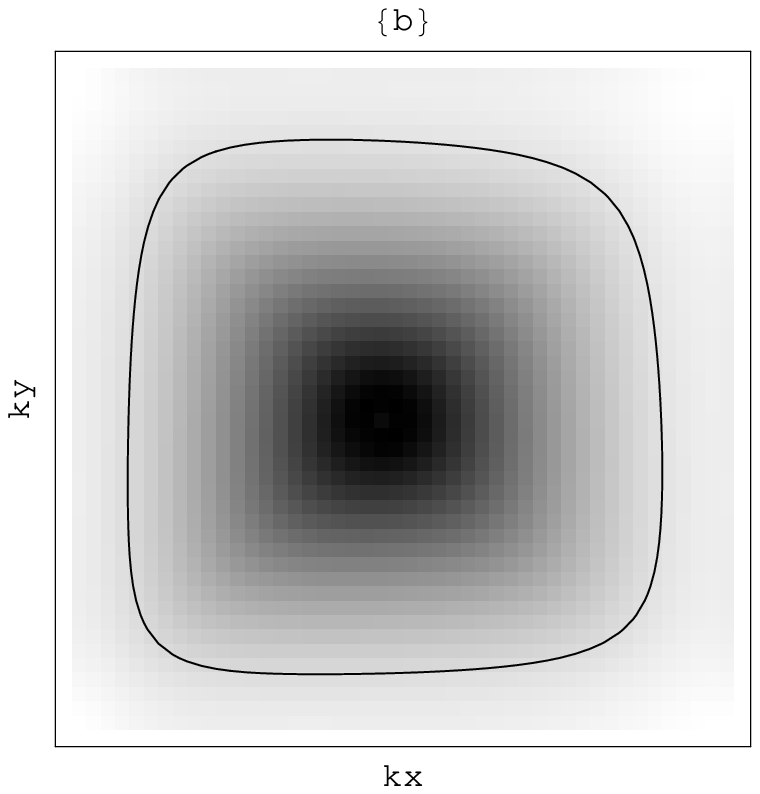}
  \caption{Energy for the two topmost bands, plotted in Grayscale,  and their respective Fermi energy contour.
  Splitting shown is due to the presence of spinorbit coupling. Notice that band (b) is shallower than band
 (a) and that the minima
  of the bands is shifted from the $\Gamma$ point reflecting the broken inversion symmetry.}
  \label{endist}
\end{figure}

\begin{figure}
  \includegraphics[height =2in]{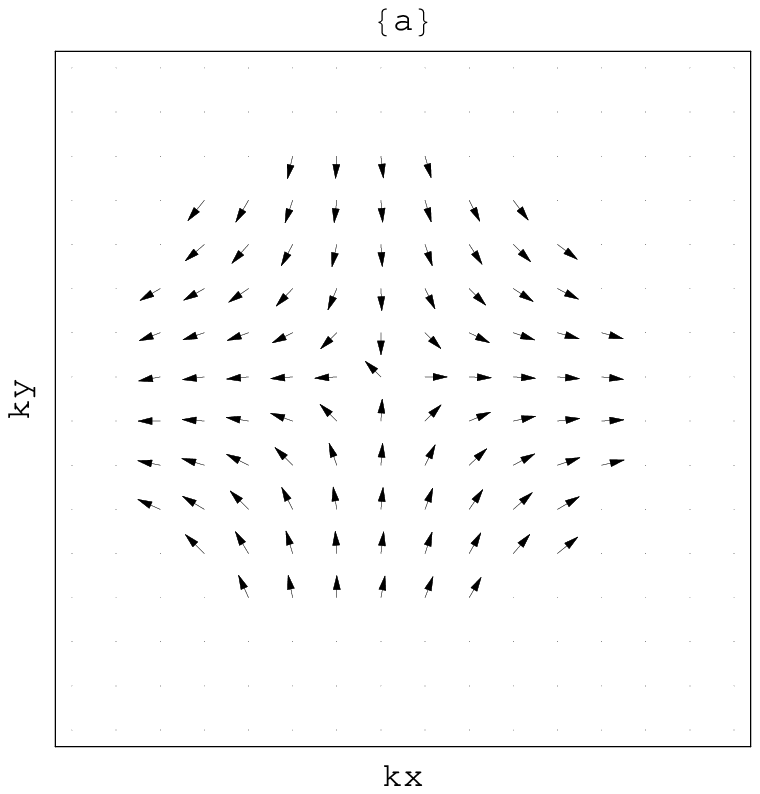}
  \includegraphics[height =2in]{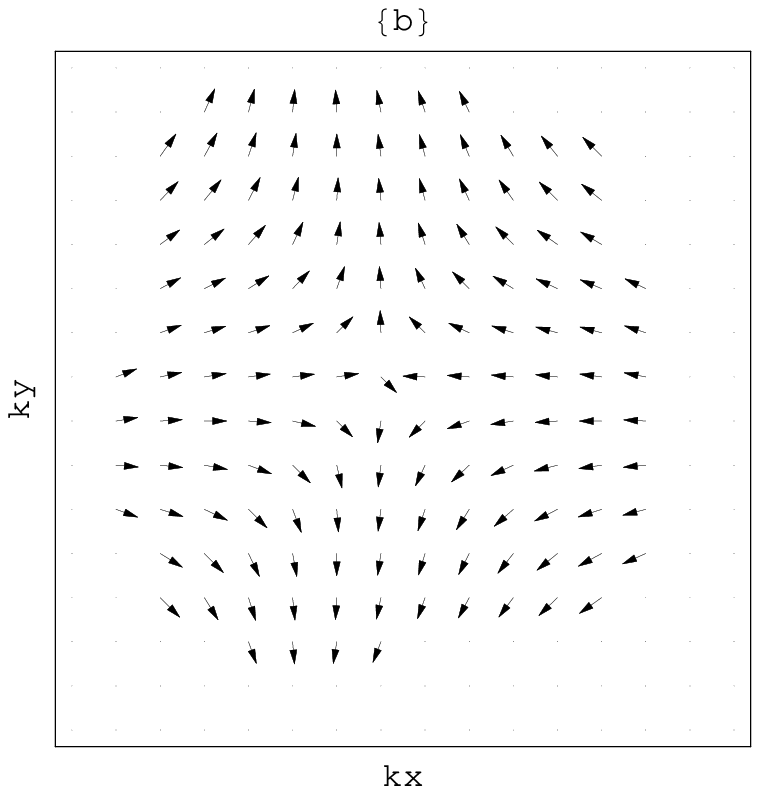}
  \caption{Orientation of spins on the Copper atom for the two topmost bands for the occupied states.}
  \label{spindist}
\end{figure}

We now  estimate numerically the direction and magnitude of the
spin-moment in the absence of exchange coupling. To do so we
discretize the Brillouin zone and for each wave vector, \textbf{k},
we find the six eigenstates and corresponding eigenvalue of
(\ref{ham}). Given the eigenstates we can compute the contribution
to the spin moment at the Copper and Oxygen sites by taking the
expectation value of their respective spin operators. We than sum
the contributions from all states below the chemical potential. For
$t_{pd} = 1$, $t_{pp} = 0.4$, $R = 0.1$, and $\lambda = 0.1$, we
find, for the domain shown in fig.\ref{cp}B, that the moment is
distributed as shown in fig.\ref{so}. To understand the origin of
the spin moment we plot in fig.\ref{endist} the energy of the two
topmost bands which are near half filling. For the occupied states
of the bands the corresponding spins on the Copper atoms are shown
in fig.\ref{spindist}. All other bands are fully filled and do not
contribute to total spin. By summing over occupied states we get the
net spin and from fig.\ref{spindist} we see that there is a net spin
along the $-\textbf{x}+\textbf{y}$ direction.

Notice, as required by our earlier symmetry argument, the spin order
breaks the $\textbf{x}=-\textbf{y}$ reflection symmetry but not the
$\textbf{x}=\textbf{y}$  reflection symmetry.
The magnetic moment due to the orbital current is estimated as $\sim
I(a^{2}/8)$ where $I$ is the current and $a$ is in-plane lattice
constant. The current density is related to the order parameter as
$(R/\bar{t}_{pd})ev_{F}a_{0}^{-3}$ where $v_{F}$ is the fermi
velocity and $a_{0}$ is the typical size of the atomic orbital. Thus
the net moment is $\sim (R/\bar{t}_{pd})(\hbar m
v_{F}/\epsilon_{F}a_{0})\mu_{B}$, where $m$ is the mass of the
electron and $\epsilon_{F}$ is the Fermi energy. For the band
structure of YBCO, $(\hbar m v_{F}/\epsilon_{F}a_{0})\sim
\mathcal{O}(1)$, we estimate the orbital moment to be $0.1\mu_{B}$ for the values of parameters chosen.
 The magnitude of the spin-moment for the same parameters is estimated from the calculations represented in Fig.~(\ref{spindist}) to be
 $0.01 \mu_B$ on the Oxygen and $0.02 \mu_B$ on the Copper.

As discussed in the section on symmetry, including exchange changes
the relative orientation of the Copper and the net Oxygen spins and might also affect the magnitude.

\subsection{Effect of Exchange Interaction on the Order of Spins}

The  microscopic
calculation above provides the result that the net spin on the two oxygens and on cu, generated due to spin-orbit coupling, are  parallel. However, once a spin-moment is
generated by the "effective field" provided by spin-orbit coupling
and orbital order, one must consider also the exchange interaction
between them to determine the direction of the moments. Exchange introduces a coupling whose leading terms are of the form

\begin{eqnarray}
H_{ex} &=&
J_{cu-o}\sum_{i}\textbf{S}^{cu}_{i}\cdot\left(\textbf{S}^{o}_{i+x}+\textbf{S}^{o}_{i-x}+\textbf{S}^{o}_{i+y}+\textbf{S}^{o}_{i-y}\right) \nonumber \\
&+& J_{cu-cu} \sum_{<i,j>}\textbf{S}^{cu}_{i} \textbf{S}^{cu}_{j}
\end{eqnarray}

\noindent where $J_{cu-o}$ and $J_{cu-cu}$  are the exchange
couplings between nearest neighbor cu and oxygen and nearest
neighbor cu's respectively. We expect the exchange interaction to be
antiferromagnetic. Hence the induced order via the effective field
discussed above competes with the the coupling $J$. In particular
one would expect a long wavelength modulation of the net moment
depending on the relative magnitude of the spin orbit interaction
and the exchange coupling. The in-plane moment is found to be of
order $10^{-2} \mu_{B}$ from which we can estimate the net effective
field $\left<\Lambda M_{O}\right> \sim \left< M_{S}\right>E_{F} =
10^{-2}$eV$/\mu_{B}$. The exchange coupling $J_{cu-o}$ is of order
$1$eV, while $J_{cu-cu}$ is of order $0.1$ev. The effective exchange
field defined as $J_{cu-o} \left<M_{S}\right>$ is then also of order
$10^{-2}$eV$/\mu_{B}$. In fig.\ref{soj} we schematically draw the
effect of exchange interaction on the in plane spins. Symmetries
dictate the existence of in plane moments while the orientation of
their ordering depends on the relative strengths of the exchange and
ordering fields. It should also be clear that the competition
between these effects will in general change the translational
symmetry of the spin-pattern. Depending on the relative exchange
parameters and the spin-orbit coupling, the spin-pattern can be very
complicated and in general incommensurate. However, since these
small moments are daughters of the orbital order, the latter is
expected to be modified only weakly. At this point, it is not
worthwhile to speculate on the details of the spin-order of the
small spin-moment of $O(10^{-2})\mu_B$ since the exchange energies
can be estimated only very imprecisely in the metallic or pseudogap
state. We must rely on the details of the magnetic structure to be
obtained  from the neutron experiments  but for such small moments,
this is no easy task.

\section{Experiments}

From the symmetry and microscopic analysis above, we have shown that
there are two sources of modulated magnetic fields within the
sample: current loops and spin order. Using polarized neutron
scattering, Fauque et al.\cite{FAQ},  performed a detailed study of
five different samples of YBCO (four underdoped and one overdoped)
to look for magnetic ordering. The most detailed measurements are on
an untwinned sample where the uncertainty is smaller.

The principle
conclusions reached were the following:

\noindent (i) A new magnetic contribution to scattering intensity
arose at the $[011]$ Bragg peak in all underdoped samples below a temperature which increases as the sample is progressively underdoped.
\noindent (ii) No signal was seen at the $[002]$ Bragg peak.

\noindent (iii) No new Bragg peaks appear ruling out breaking of
translation invariance.

\noindent (iv) The data could be fit to the current loop model
provided one assumed that the moments were located at the centroid
of the triangles.

\noindent (v) Assuming that all the signal was due to a single
source of magnetic ordering, i.e. it has a unique form-factor and structure
factors, the moment had to be tilted away from the c-axis. The angle
was largest  for the detwinned sample with the moment being at $\sim
45^{\circ} \pm 20^{\circ}$. Our finding that the in-plane moment is due to spins while the out of plane component is due orbital order necessitates a reevaluation of these numbers.

To fit the neutron scattering data one has to assume a model for the
magnetic moments in the system. From the observations above we
conclude that within the experimental uncertainties,  $(\sim 0.01 \mu_{B})$,  the moment is commensurate with the lattice and that
the net moment in a copper oxide plane is zero. The latter follows from absence of observable magnetic signal at $[002]$.  For loop-current order, the spin
flip signal should appear at Bragg Peaks $[0,K,L]$,
$[H,0,L]$, $[H,H,L]$ and $[H,-H,L]$. The dimpling of the plane
implies that these moments are at an angle of $\sim 7^{\circ}$ with
the c-axis. To understand the origin of the larger deduced angle as
stated in point (v) above we have to take into account the fact
that the in-plane moments arise due to ordering of spins while the
out of plane component is due to current loops. The
corresponding moments have very different form and structure
factors. For any given ordering of moments $\textbf{M}(\textbf{r})$
the spin flip scattering intensity at Bragg peak $\textbf{Q}$ =
$[H,K,L]$ for polarization of the incident Neutron parallel to
$\textbf{Q}$ is \cite{ML}

\begin{eqnarray}
I\left(\textbf{q}\right) &\propto &
\left|f_{Q}\right|^{2}\left|S(\textbf{Q})\right|^{2}\left|M_{\perp}\right|^{2}\\
\nonumber \textbf{M}_{\perp} &=&
\textbf{Q}\times\left(\textbf{M}\times \textbf{Q}\right)/Q^{2}
\end{eqnarray}

\noindent where $f_{Q}$ is the form factor, given by the fourier
transform of the spread of individual moment,
$S\left(\textbf{Q}\right)$ is the structure factor, given by the
fourier transform of the distribution of these localized moments in
the crystal and $M$ is the magnitude of the localized moment. For
the current loops, the magnitude of the moments is of order $0.1
\mu_{B}$, while for the spins it is $0.02 \mu_{B}$. The structure
factor for the current loops is proportional to $\cos\left(\pi z
L\right)\sin\left(2\pi x_{0}\left(H\pm K\right)\right)$, where $z
\sim 0.29$ is the ratio of the interlayer spacing to the lattice
constant $c$ and $x_{0}$ is the position of the centroid of the
triangular plaquette given by $\left(x_{0},x_{0}\right)$. The cosine
factor reflects the fact that the orbital moments are identical in
the bilayers while the sine factor arises due to the
antiferromagnetic orientation between the two triangles in the unit
cell. For the spins, the in-plane structure factor cannot be completely
determined without knowing the precise relative angles of the spins
in a unit-cell. But
 the fact that the spins are oppositely oriented in the two
layers implies that the structure factor can be as large as
$\sin\left(\pi z L\right)$. Thus at $[011]$ the spin structure
factor can be as large as $1.8$ times the moment structure factor.

For current loops, the effective moment generated is spread over the
area of the triangular plaquette which is $a^{2}/8$, where, $a$ is
the lattice constant. For spins on copper and oxygen atoms, the
moment is distributed over the $d_{x^{2}-y^{2}}$ and $p_{x,y}$
orbitals respectively. Since the atomic orbitals are more localized
their fourier transforms are weaker functions of $\textbf{Q}$ as
compared to the orbital moments. To estimate the form factors we
model the time reversal violating state with current wires along the
$x$, $y$ and $-x-y$ directions with thickness $\delta$. The fourier
transform of this pattern of currents is expressed in terms of a combinations of
form and structure factors. Then the form factor is
$2\exp\left(-\pi^{2}\delta^{2}/a^{2}\right)/\pi$ where $a$ is the
lattice constant. We have assume a Gaussian profile for the current
in the wires. Since the width of the current wire is related to the
overlap of the copper and oxygen orbitals, we take it to be of order
$1 \AA$. Thus the form factor for the current loop is $\sim 0.3$.
The form factor for the spins is $\sim 0.9$ implying that the net
geometric factor for spins is $\sim 6$ times larger than those for
the current loops. Give the estimate for the magnitude of the spin
and orbital moments, the rough estimate for the geometric factors
implies that indeed the resulting neutron scattering intensities due
to the two orderings will be of the same order of magnitude.

In conclusion, we find that given the orbital order, an inplane
spin-order is mandated. With reasonable assumptions about the
relative form factors for loop-currents and spin-moments and the
calculated magnitude of the ordered spin-moment, the observed
polarized diffraction can be understood. A detailed test awaits
experimental refinements. Given the spin structure shown in fig.3,
DM interactions can induce a tilting of the spins leading to a small
ferromagnetic moment. This is also under further investigation.

 {\it Acknowledgements}: We wish to acknowledge invaluable
discussions with Philippe Bourges on details of the experimental
results and their interpretation.
\bibliography{at}% Produces the bibliography via BibTeX
\bibliographystyle{unsrt}

\end{document}